\documentclass[article,twocolumn,english,secnumarabic,footinbib,tightenlines,nobibnotes,aps,prl,unsortedaddress,superscriptaddress,showpacs]{revtex4-1}
\usepackage[usenames,dvipsnames]{color}
\usepackage{amsmath}
\usepackage{amsfonts}
\usepackage{amssymb}
\usepackage{mathtools}
\usepackage{pict2e}
\usepackage{hyperref}
\usepackage{graphicx,xcolor}
\usepackage{bbding}
\graphicspath{ {./Figs/} }
\usepackage{bm}
\usepackage{multirow}
\usepackage{bm}
\usepackage{multirow}

\usepackage{grffile}

\usepackage[caption=false]{subfig}

\newcommand{\avbra}[1]{\blangle #1 \brangle}

\DeclareMathOperator{\Imag}{Im}
\DeclareMathOperator{\Real}{Re}
\newcommand{\mib}{\bm}
\newcommand{\beq}{\begin{equation}}
\newcommand{\eeq}{\end{equation}}
\newcommand{\be}{\begin{equation}}
\newcommand{\ee}{\end{equation}}
\newcommand{\bea}{\begin{eqnarray}}
\newcommand{\eea}{\end{eqnarray}}
\newcommand{\ba}{\begin{array}{ccc}}
\newcommand{\ea}{\end{array}}

\newcommand{\cbra}[1]{\left \{ #1 \right\}}

\newcommand{\drho}{\delta \rho}

\mathchardef\sPhi="7108
\mathchardef\sLambda="7103
\mathchardef\sGamma="7100
\mathchardef\sDelta="7101
\mathchardef\sOmega="710A
\def\ssr#1{{\sss{\rm #1}}}
\def\frac#1#2{{\textstyle{#1 \over #2}}}
\def\half{\frac{1}{2}}
\def\fourth{\frac{1}{4}}
\def\blangle{{\big\langle}}
\def\brangle{{\big\rangle}}
\def\sket#1{{|  \, #1 \,  \rangle}}

\def\ctnh{{\rm ctnh}}

\def\Bb{{\mib b}}
\def\Br{{\mib r}}
\def\Bx{{\mib x}}
\def\Be{{\mib e}}
\def\Bk{{\mib k}}
\def\Bq{{\mib q}}
\def\BG{{\mib G}}
\def\BR{{\mib R}}
\def\BQ{{\mib Q}}
\def\Btheta{{\mib\theta}}
\def\Bxi{{\mib\xi}}
\def\HR{{\hat R}}
\def\RK{{\rm K}}

\def\RN{{\rm N}}
\def\RH{{\rm H}}
\def\RP{{\rm P}}
\def\ctil{{\tilde c}}

\def\ns{^{\vphantom{*}}}

\def\pz{\partial}
\def\sss#1{{\scriptscriptstyle #1}}
\def\ssr#1{{\sss{\rm #1}}}

\def\bvph{\vphantom{\int\limits_A^B}}
\def\HOmega{{\hat\sOmega}}
\def\impi{\int\limits_{-\infty}^{\infty}\!\!}

\def\kT{k\ns_\ssr{B}T}

\begin{document}

\title{Collective modes in a quantum solid}

\author{Snir Gazit}
\affiliation{Department of Physics, University of California, Berkeley, CA 94720, USA}
\affiliation{Physics Department, Technion, 32000 Haifa, Israel}
\author{Daniel Podolsky}
\author{Heloise Nonne}
\author{Assa Auerbach}
\affiliation{Physics Department, Technion, 32000 Haifa, Israel}

\author{Daniel P. Arovas}
\affiliation{Department of Physics, University of California at San Diego, La Jolla, California 92093, USA}

\date{\today }

\begin{abstract}
We provide a theoretical explanation for the optical modes observed  in inelastic neutron scattering (INS) on the bcc solid phase of helium 4  [T. Markovich, E. Polturak, J. Bossy, and E. Farhi, Phys.~Rev.~Lett.~88, 195301 (2002)].  We argue that these excitations are amplitude (Higgs) modes associated with fluctuations of the crystal order parameter  {\em within} the unit cell. We present an analysis of the modes based on an effective Ginzburg-Landau model, classify them according to their symmetry properties, and compute their signature in INS experiments. In addition, we calculate the dynamical structure factor by means of an {\em ab intio} quantum Monte Carlo simulation and find a finite frequency excitation at zero relative momentum.  
\end{abstract}

\maketitle

{\em Introduction --}  The harmonic theory of solids predicts that the excitation spectrum of a monoatomic Bravais lattice crystal should consist solely of acoustic phonons.  It is
therefore quite surprising that inelastic neutron scattering measurements have detected gapped optic-like modes in the bcc phase of solid helium 4 \cite{Markovich,pellegPRB,PellegJLTP}. 
These gapped excitations include a dispersing longitudinal mode \cite{Markovich} and a non-dispersing transverse mode \cite{pellegPRB,pellegThesis,PellegJLTP}.  Similar measurements
carried out in the hcp phase of $^4$He have tentatively identified optic-like modes {\em beyond} those expected for the hcp structure \cite{blackburn2008roton}.
Previous theories associated these excitations with non-phononic dipolar  \cite{Gov} and delocalized vacancy modes \cite{blackburn2008roton}. While these scenarios are plausible, 
they do not explain why such modes are unique to $^4$He, nor can they account for the existence of multiple dispersing gapped modes.  Hence, a basic understanding of this
phenomenon is still lacking.

In this Letter, we identify these optical excitations as amplitude modes of the solid order parameter.   Amplitude (``Higgs'') modes  appear in many condensed matter systems and have recently been the focus of intense experimental and theoretical research. Some notable examples are superconductors \cite{LittlewoodVarma,FrydmanNature,Podolsky_visibility}, cold atoms in an optical lattice \cite{Huber_Amplitude,endres},  and antiferromagnets \cite{ruegg}.  However, such modes are not generally expected in solids.  Here we argue that the large zero point fluctuations in solid $^4$He allow it to be treated as a three dimensional charge density wave (CDW) that supports, in addition to the usual acoustic phonons, gapped amplitude modes.

The phonon spectrum of bcc ${}^4$He is strongly dependent on quantum effects, and for two reasons.  First, the classical bcc structure is difficult to stabilize with
power law potentials, and for Lennard-Jones
systems, the only equilibrium states are either hcp or fcc \cite{Travesset2014}, at all molar volumes.  Second, the small helium mass leads to large zero-point fluctuations on the order of
30\% of the interatomic distance \cite{glyde1994}, well in excess of the Lindemann criterion for thermal melting.  This invalidates the use of real space atomic displacements as
small expansion parameters, and entails significant nonlinear contributions to the equations of motion.  To account for quantum fluctuations, Hartree \cite{PhysRev.162.824} and self-consistent
harmonic theories \cite{PhysRevA.5.2230,Guyer1970,Glyde1971,Horner1972} have been derived, and have met with general success in describing the experimentally observed acoustic phonon spectra
\cite{PhysRevA.5.1537,PhysRevA.8.1513}.  If strong enough, the nonlinear quantum lattice dynamics can in principle support excitations beyond the acoustic phonons.  Physically, the large
zero point motion may allow for density fluctuations {\em within} the unit cell, which are likely responsible for the optical modes seen in neutron scattering.

\begin{figure}[b]
\includegraphics[scale=0.45]{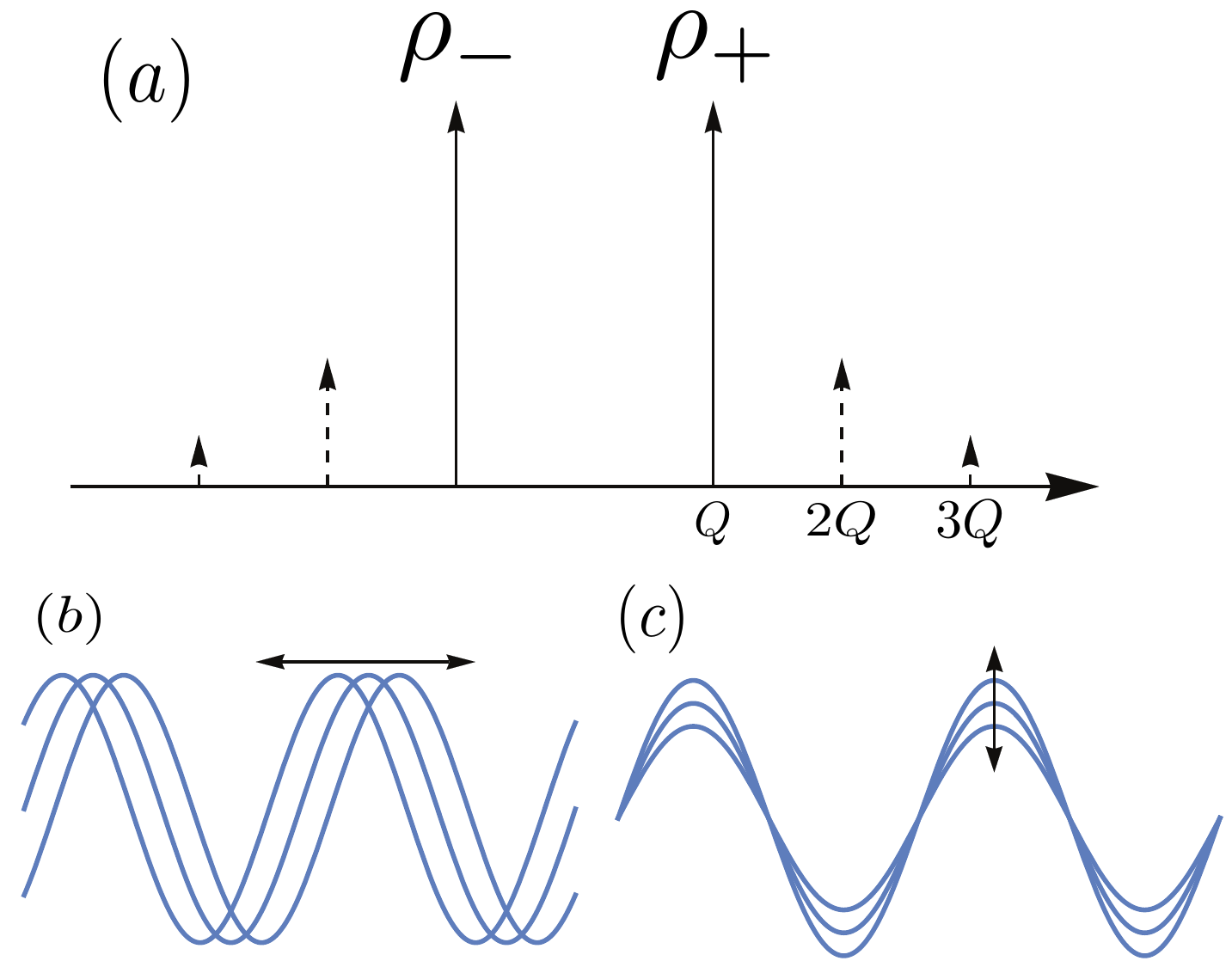}
\centering
\caption{{\bf (a)}  A one-dimensional CDW  is modulated at primary Bragg vectors $\pm Q$, with Fourier amplitudes $\rho_\pm$.
Higher harmonics are strongly suppressed.  Real space oscillations of: {\bf (b)} the acoustic phonon and {\bf (c)} the amplitude mode.}
\label{fig:fig1}
\end{figure}

A natural question is then how to construct a linear theory for the excitations of bcc $^4$He. To address this problem, we note that the large spread of the atomic density profile within the unit cell leads to suppression of high lattice harmonics. This is in sharp contrast with the classical picture of a crystal with well localized atoms, for which {\em all} lattice harmonics are sizable.  It may therefore be justified to neglect high harmonics  and consider only the Fourier weight corresponding to the primary Bragg vectors. The resulting description of the solid is then in terms of a CDW, whose spectrum naturally contains both acoustic phonons and amplitude modes (see Fig. \ref{fig:fig1}).

{\em Ginzburg-Landau theory -- } In constructing the phenomenological Ginzburg-Landau theory   we make two main assumptions.  First, the gapped modes are presumed to result from density fluctuations (and not, {\em e.g.}, from fluctuations related to nearby superfluid order).  Second, we assume the density dynamics is dominated by a small number of primary Bragg vectors. Our results follow from these assumptions combined with symmetry considerations.  Hence we expect them to be robust and apply beyond the specific choice of the phenomenological model used. 

For a solid with large density fluctuations, the GL free energy functional can be written as an expansion in powers of the deviation of the particle density from its average value, $\drho(r)=\rho(r)-\rho_0$ \cite{BaymBethePethick,Alexander_bcc_GL},
\begin{eqnarray}
\mathcal{F}_{\text{GL}} [\drho(\Br)] & = & \int \!\! d^3 r\> d^3 r' \,\drho(\Br')  \,\chi ^{-1}(\Br-\Br') \, \drho(\Br) \\
                && - B \! \int \!\! d^3 r \, \left[\drho(\Br)\right]^3+C\!  \int \!\! d^3r \,  \left[\drho(\Br)\right]^4\,. \nonumber
\label{eq:GL_static}
\end{eqnarray}
The charge susceptibility kernel $\chi(\Br-\Br')$ must respect the underlying rotational symmetry and support an instability towards a CDW at a Bragg vector of magnitude $G$. These considerations are fulfilled by taking 
\begin{equation}
\chi^{-1}(\Br-\Br')=\frac{1}{2}\!\left[R+v^2 (\nabla^2+G^2)^2\right]\delta(\Br-\Br')
\end{equation}
  for which the quadratic term in Eq.~\eqref{eq:GL_static} is minimized for all wavevectors $\BQ$ of length $G$.  For $R<0$, any Fourier component of the density with $|\BQ|=G$ contributes a negative free energy, thus leading to a CDW instability.  The wavevector pattern $\{\BQ\ns_i\}$ selected is then determined by higher order terms in $\mathcal{F}_{\text{GL}}$. 

As shown in Refs. \cite{BaymBethePethick,Alexander_bcc_GL}, the cubic term in ${\mathcal{F}_{\text{GL}}}$ prefers structures which maximize the number of equilateral triangles
which can be formed from the $\{\BQ\ns_i\}$.  For three-dimensional crystals, this selects fcc in reciprocal space, hence bcc in real space.
The density profile of the CDW is then
$\drho\ns_{\scriptscriptstyle{\rm CDW}}(r)=\sum_{\BG} \rho\ns_{\BG} \, e^{i\BG\cdot \Br}$,
where the sum runs over the twelve primary Bragg vectors of the bcc lattice.   Since $\rho(r)$ is real, we must have $\rho\ns_\BG=\rho^*_{-\BG}$,
hence the total number of real degrees of freedom is twelve, as opposed to three in the harmonic lattice theory.   For the minimum energy configuration, $\rho\ns_\BG$
can be chosen to take a uniform ($\BG$-independent) mean field value $\rho\ns_\BG=\delta\bar{\rho}$.   Excitations are then obtained by studying dynamical fluctuations
of the density about the mean-field solution.

{\em Excitation spectrum  --} 
In order to study excitations, we consider the time-dependent GL Lagrangian density
\begin{eqnarray}
\mathcal{L}={1\over \gamma}\bigg({\partial\,\delta\rho(\Br,t)\over\partial t}\bigg)^{\!\!2} -\mathcal{F}_\text{GL}[\drho(\Br,t)]\ ,
\label{eq:GL_dynamic}
\end{eqnarray}
where $\mathcal{F}_\text{GL}$ is the free energy defined in Eq.~\eqref{eq:GL_static}. Note that since the density is real, $(\partial\ns_t\rho)^2$ is the lowest order dynamical term that can be constructed. In principle, dissipative terms involving first order time derivatives are also allowed.  These terms originate from processes in which gapped modes decay to acoustic phonons and lead to broadening of the line shapes.  Experimentally, the modes are found to be sharp and hence we will neglect this effect in our phenomenological approach.  

\begin{figure}[t!]
\centering
  \subfloat[Longitudinal]{\includegraphics[width=0.25\textwidth]{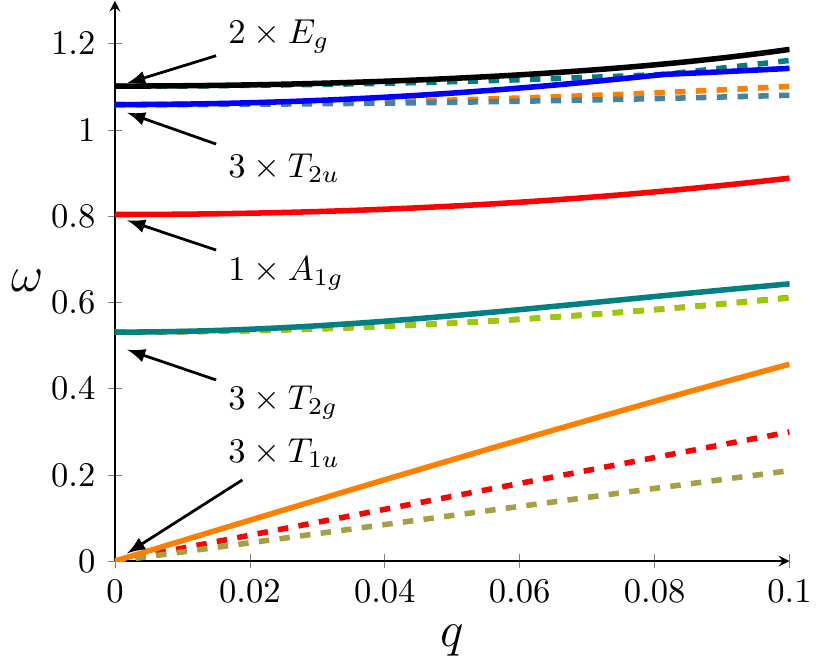}\label{exc_lon}}
  \subfloat[Transverse]{\includegraphics[width=0.25\textwidth]{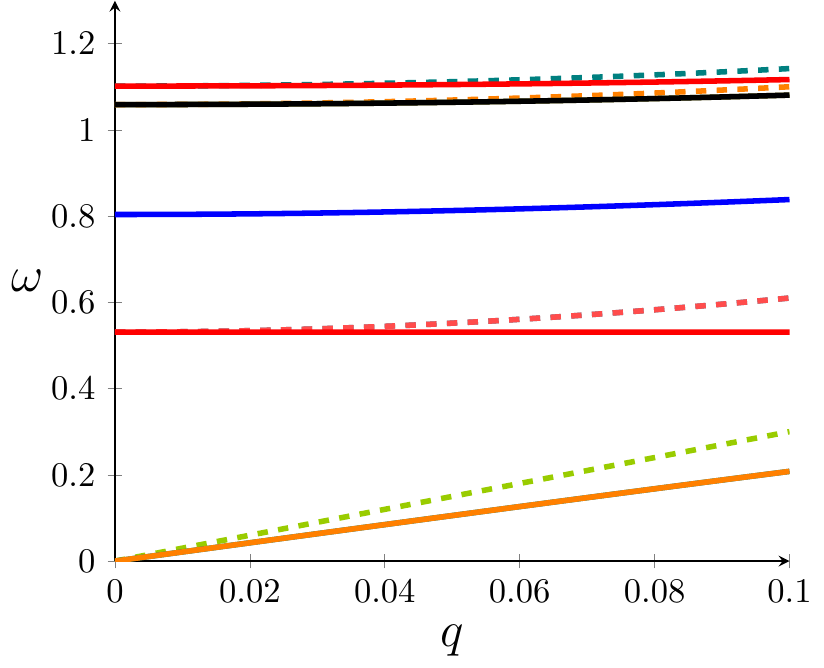}\label{exc_tra}}
\caption{Excitation spectrum for the GL theory of BCC $^4$He. GL parameters are chosen as $R=0.05,B=C=1$ and $v=1.5$.  The momentum transfer $\Bq$ is chosen relative to the principal Bragg vector $G=\{1,1,0\}$
in the  (a) longitudinal  $\Bq=(q,q,0)$ and (b) transverse $\Bq=(0,0,q)$ directions. Solid (dashed) lines correspond to INS active (inactive) modes. Modes labelled in
panel (a) by their irreps and mode degeneracies at $q=0$.}
\label{exc-spec}
\end{figure}

The excitation spectrum is obtained by linearizing the Euler-Lagrange equations\ with respect to the fluctuations
$\eta\ns_\BG(q,\omega)=\rho(\BG+q,\omega)-\delta \bar{\rho}$. Diagonalizing the resulting bilinear form yields the linear mode eigenfrequencies $\omega_\alpha$ and
eigenvectors $\Bxi_\alpha$, as discussed in the Supplementary Material (SM).  The longitudinal and transverse excitation spectrum in the vicinity of a principal Bragg vector is shown in Figs. \ref{exc_lon} and \ref{exc_tra}.  In addition to the three acoustic phonon branches the spectrum also contains nine gapped optical modes.

{\em Symmetry classification --} At zero relative momentum, $q=0$, the normal modes can be classified according to irreducible representations (irreps) of the octohedral group $O_h$. In Fig. \ref{exc_lon} we label the different modes according to their irreps. We find that the two lowest gapped modes are one $s$-wave and three $d$-wave $T_{2g}$ modes.  When the GL parameter $R$ is below a critical value $R^*$, the d-wave mode is lower in energy than the s-wave modes; this order is exchanged for $R>R^*$.  This is the only qualitative feature of our analysis that depends on the precise value of the GL parameters.

\begin{figure}[t!]
\centering
  \subfloat[$s$-wave]{\includegraphics[width=0.20\textwidth]{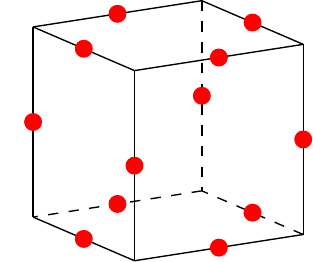}}
  \subfloat[$d$-wave]{\includegraphics[width=0.20\textwidth]{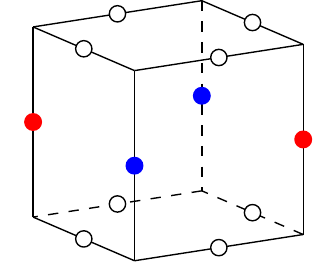}}\\
  
  \vskip 4mm  
  
  \subfloat[$d_{xy}$-wave time evolution]{\includegraphics[width=0.48\textwidth, trim=0 20 4.5 0,clip]{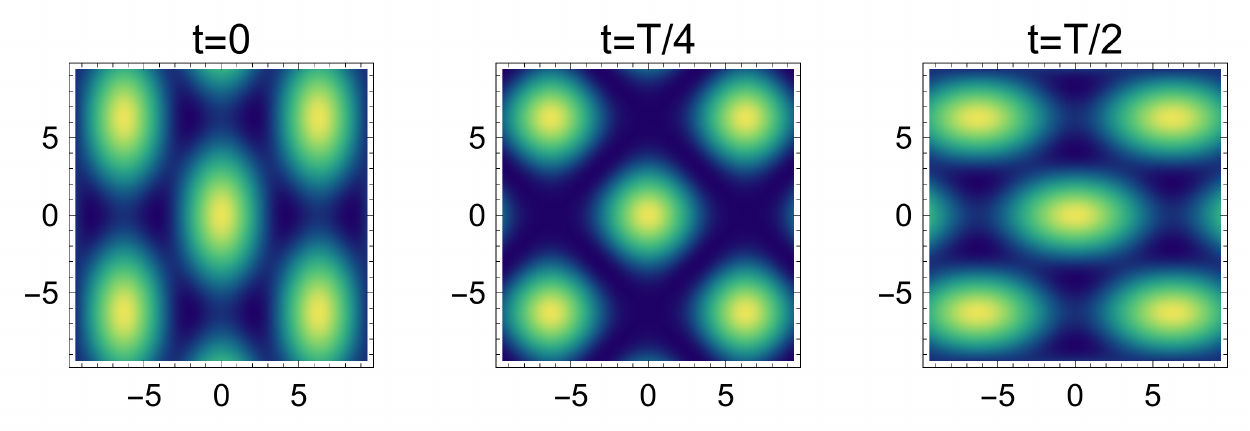}\label{dwave_anim}}

\caption{Reciprocal space structure of the (a) s-wave and (b) d-wave normal modes at the primitive Bragg vector. Red (blue) dots correspond to positive (negative) amplitudes $\eta_\BG$.  (c) Real space  evolution of the density for the $d_{xy}$ mode at $q=0$, at different instants in time.}
\label{fig:modesymmetry}
\end{figure}

Interestingly, the $d_{xy}$ mode is approximately dispersionless in the transverse direction, $\Bq=q\hat{{\mib z}}$, as seen in Fig. \ref{exc_tra}. This behavior, seen experimentally \cite{pellegPRB,PellegJLTP}, can be traced back to the reciprocal space structure of the mode depicted in Fig. \ref{fig:modesymmetry}, which has non vanishing amplitude $\eta\ns_\BG$ only for $\BG$ in  the $z=0$ plane. As a result, $\BG\cdot\Bq=0$, and the charge susceptibility kernel $\tilde{\chi}^{-1}(\BG+\Bq)=\half R+\half v^2((\BG+\Bq)^2-G^2)^2=\half(R+v^2 q^4)$ has a weak dependence on the relative momentum $q$, leading to an excitation that disperses only as $q^4$. 

In Fig. \ref{dwave_anim} we plot the real space density profile corresponding to the $d_{xy}$ mode at $q=0$. The images are projected onto the $z=0$ plane and track the time evolution at quarter and half of the oscillation period $T$.  Note that the oscillations are quadrupolar deformations of the density {\em within} the unit cell such that the mode can support a finite frequency oscillation at zero momentum. This behavior is analogous to optical phonons in polyatomic crystals, which involve relative motion of the atoms within a unit cell. 

{\em Dynamical structure factor  --}
In order to compare with the experimental inelastic neutron scattering (INS) data we compute the dynamical structure factor (DSF),
$S\ns_\BG(\Bq,\omega)=\Imag\left\langle\delta \rho\ns_{\BG}(\Bq,\omega)\,\delta \rho\ns_{-\BG}(-\Bq,-\omega)\right\rangle$.
This is readily computed via canonical quantization of the normal modes (see SM), yielding
\begin{equation}
\begin{split}
S(\BG+\Bq,\omega)=&\sum_{\alpha}{\gamma M_{\BG,\alpha}(\Bq)\over 2\omega_{\alpha}(\Bq)}\Bigg\{n(\omega_\alpha)\,\delta\big(\omega+\omega_\alpha(\Bq)\big)\\
&\quad+\ \big(1+n(\omega_\alpha)\big)\,\delta\big(\omega-\omega_\alpha(\Bq)\big)\Bigg\}\ ,
\label{eq:structure_factor}
\end{split}
\end{equation}
where $n(\epsilon)=1\big/[\exp(\beta \epsilon)-1]$ is the Bose function, and the matrix element $M_{\BG,\alpha}(\Bq)$  expresses the overlap squared between normal mode $\alpha$ and the density operator. For a given wavevector, only certain normal modes are INS active, while the remaining modes have vanishing $M_{\BG,\alpha}(\Bq)$. This is illustrated in Fig.~\ref{exc-spec}, where INS active (inactive) modes are plotted with solid  (dashed) lines. As a check, the DSF analysis predicts that the longitudinal (transverse) acoustic phonon are INS inactive for $\Bq$ vector which are orthogonal (parallel) to the Bragg vector $\BG$. This reproduces well-known selection rules of the classical harmonic theory of solids \cite{marder2010condensed}. In addition, our analysis uncovers new selection rules that apply to the gapped modes.  

The DSF in Eq.~\eqref{eq:structure_factor} satisfies the sum rule \cite{sum_rule},
\begin{eqnarray}
\int\limits_{-\infty}^{\infty}\!\!\!d\omega\> \omega \,S(\BG+\Bq,\omega)=\gamma\, .
\end{eqnarray}
Hence,  although our model contains more phonon modes than predicted by the harmonic theory of solids, their total spectral weight is constrained.  In particular, the distinction between a crystal and a CDW is qualitative in nature and, as such, there must be a smooth mechanism by which the amplitude modes disappear as the solid becomes more classical.  Indeed, in this limit, the amplitude modes become increasingly energetic, and their spectral weight is accordingly reduced in order to satisfy the sum rule.  As a related observation, if we were to enlarge our GL formalism by including non-primary Bragg vectors, we would obtain more optical modes, but of very high energy and small spectral weight.

{\em Quantum Monte Carlo -- } 
We complement the phenomenological GL analysis with an  {\em ab initio} path integral quantum Monte Carlo (QMC) simulation. We model the $^4$He atoms with the following Hamiltonian,
\begin{equation}
H=-\lambda \sum_{i=1}^{N} \vec{\nabla}_i^2 +\sum_{i<j} V(r_i-r_j),
\end{equation}
where $\lambda=6.0596\,\text{\AA}^{\!2}\,\text{K}$ for $^4$He and $V(r)$ is the Aziz potential, which is believed to accurately capture the inter-atomic
potential energy of $^4$He atoms \cite{aziz}. We set the simulation parameters to lie within the bcc region of the $^4$He phase diagram. Explicitly, the
temperature is set to $T=1.6$ K and we consider $N=2000$ $^4$He atoms at atomic density $n_0=0.02854\,\text{\AA}^{-3}$ (molar volume
$v\ns_0=21.1\,{\rm cm}^3$). The bosonic world-lines configurations are sampled employing the continuous space worm algorithm \cite{BoninsegniWorm}
in the canonical ensemble.  Our main observable is the charge susceptibility structure factor evaluated at Matsubara frequency $\omega_m$,
\begin{equation}
\chi(\Bq,i\omega_m)={1\over N \beta}\left\langle\>\left|\int\limits_0^\beta\! d{\tau}\>e^{i\omega_m\tau}\sum_{i=1}^{N} e^{i \Bq\cdot \Br_i(\tau)}\right|^2\>\right\rangle.
\label{qmc_sf}
\end{equation} 
where $\Br_i(\tau)$ denotes the position of the $i^{\rm th}$ particle at imaginary time $\tau$.  

At finite relative momentum the spectrum is gapped up to the energy scale of the acoustic phonon and hence the dispersion relation can be extracted by fitting the DSF
to the asymptotic form $\chi(q,\tau\gg 1/\Delta) \sim f(q,\tau)$ where $f(q,\tau)=A(q) (e^{-\tau \Delta(q)} +e^{-(\beta-\tau) \Delta(q)}) $. In addition, we compute the
excitation spectrum by performing a numerical analytic continuation on the imaginary time QMC data using the MaxEnt method \cite{analytic_cont}. We find good agreement between the excitation spectra computed via these two methods.

\begin{figure}
\includegraphics[scale=0.9,trim={10.5cm 0 9cm 0},clip]{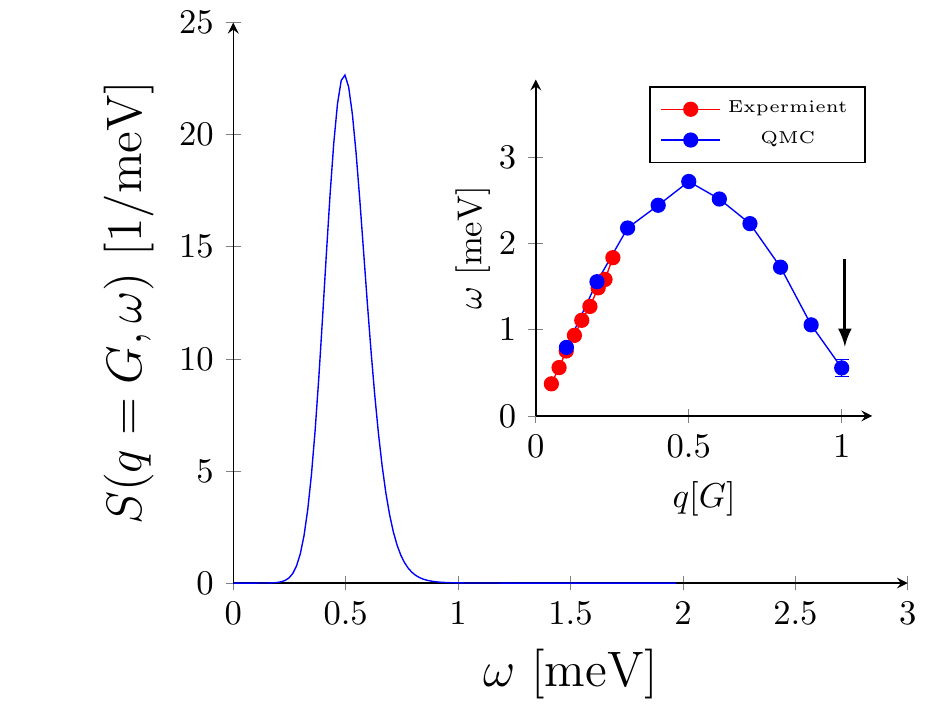}
\centering
\caption{Dynamical structure factor at the primitive Bragg vector $\BG=\{1,1,0\}$. The spectral function displays a clear resonance at $\omega=0.5(1)$ meV that we attribute to fluctuation of the CDW order parameter. The first moment $I_1=\int\! d\omega\, \omega\, S(\omega,\BG)$ obeys the $f$-sum rule $I_1^{\text{e}}={\BG^2 /2m} =2.43$ meV. The numerical value, $I_1^{\text{n}}=2.6$ meV, deviates by $7\%$ from the exact result. {\sl Inset:} Dispersion of lowest energy longitudinal excitation computed with QMC simulations (blue), compared to longitudinal acoustic phonon measured in INS experiments (red) \cite{Markovich}.  The arrow indicates the optical mode. }
\label{fig:qmc}
\end{figure}

We focus on the longitudinal mode along the line connecting the Brillouin zone origin to a primitive Bragg vector.  The resulting dispersion relation is depicted in
the inset of Fig. \ref{fig:qmc}. For comparison we also display the experimental INS data \cite{Markovich} and find good agreement with the numerical computation.  
We were unable to resolve any gapped modes at finite relative momentum. 
This is likely due to the small spectral weight of the optical mode relative to the acoustic phonons. 
Experimentally, the ratio is about 1/10. A previous QMC study \cite{QMCbccHe} computed the acoustic dispersion at finite wave vectors, and
thus  did not  detect the optical modes.

To overcome this problem we compute the DSF at a Bragg momentum, where the weight of the acoustic phonons is expected to vanish. To extract the energy scale of
the amplitude mode we fit the imaginary time QMC data to the following form $\chi(\BG,\tau\gg 1/\Delta) \sim \chi_0+ f(\BG,\tau)$. The extensive constant $\chi_0$
takes into account the crystal order parameter. A typical fit of this type is shown in the SM. As before, we also perform numerical analytic continuation of the QMC data,
and obtain consistent results. 

The results of  the numerical analytical continuation are presented in the main panel of Fig. \ref{fig:qmc}. We find the spectrum is composed of a gapped resonance
peaked at $\omega\ns_{\rm H}= 0.5(1)$ meV.  This value is smaller than the experimentally measured frequency $\omega\ns_{\rm H}=1.2$ meV.  The disagreement
might be due to numerical errors originating from the numerical analytic continuation, or systematic errors in modeling the interatomic potential.  A more interesting
possibility is that this low frequency mode exists, but was not observed in INS due to experimental limitations \cite{private_emil}. We have, nevertheless,
demonstrated numerically the presence of a gapped mode at zero relative momentum in a monoatomic Bravais lattice. 

Could the experimentally observed optical mode be due to a {\it two\/} phonon process \cite{Bert}?  In the SM, we consider, analyze, and reject this possibility. 
We find that the peak in the two-phonon contribution to $S(\BG,\omega)$ should lie at a frequency of $\omega\ns_{\rm 2-ph}\simeq 4.3\,$meV, and furthermore the
peak amplitude of the two-phonon contribution to $S(\BG,\omega)$ is $0.6\%$ of the peak value in Fig. \ref{fig:qmc}.

{\em Discussion -- } Our results motivate future experimental studies of the excitation spectrum of solid $^4$He. Specifically, it would be interesting
to determine the symmetry properties of the gapped modes. Breaking a sub group of the $O_h$ crystal symmetry group, {\it e.g.\/} by shearing or compressing the
lattice would lift the symmetry enforced degeneracy. The resulting splitting of the phonon branches could be detected in INS experiments.

More broadly, beyond $^4$He, our analysis may be relevant to other examples of strongly fluctuating quantum solids.   In that regard, one promising future theoretical and experimental research direction would be to explore the effect of reduced dimensionally on the amplitude modes such as in the solid phase of two dimensional dipolar Bose gases \cite{dipolarGas}.   As a more concrete prediction, the Lindemann parameter for bcc solid $^3$He is even larger than that of $^4$He \cite{glyde1994}, and therefore we predict that lower energy optical modes should be seen in INS measurements in solid $^3$He.  This would also serve as an experimental confirmation that the optical modes in $^4$He are strictly due to charge fluctuations and not to gapped fluctuations of the nearby superfluid.

Summarizing, we have identified the gapped modes observed in INS experiments on the bcc phase of solid $^4$He with amplitude fluctuations of the crystal order. The properties of the gapped modes were analyzed though an effective GL theory and an {\em ab-inito} QMC simulation. In addition, we propose experimental tests for our predictions in solid $^4$He and quantum solids in general. 

{\it Acknowledgements -- } We thank B.~I.~Halperin, J.~Goodkind, E.~Polturak,  and S.~Sinha for helpful discussions.  AA and DPA gratefully acknowledge support from the
US-Israel Binational Science Foundation grant 2012233. AA and DP acknowledge  support from the Israel Science Foundation and thank the Aspen Center for Physics,  supported by the NSF-PHY-1066293, for its hospitality

\bibliography{Helium}{}

\begin{thebibliography}{32}
\bibitem [{sum()}]{sum_rule}%
  \BibitemOpen
  \href@noop {} {}\bibinfo {note} {The familiar $f$-sum rule is $\int d\omega\>
  \omega \,S(\Bk,\omega)=\Bk^2/2m$, where $m$ is the particle mass. For
  $\Bk=\BG+\Bq$ with $q\ll G$, the RHS of the sum rule is $G^2/2m + {\cal
  O}(q)$, and is approximately a constant, which we can use to fix the
  coefficient $\gamma$ of the time derivative in our Lagrangian
  density.}\BibitemShut {Stop}%
\end{thebibliography}

\newpage
\onecolumngrid

\begin{center}
	\textbf{\large Supplementary material for ``Collective modes in a quantum solid"}
\end{center}
\setcounter{equation}{0}
\setcounter{figure}{0}
\setcounter{table}{0}
\makeatletter
\renewcommand{\theequation}{S\arabic{equation}}
\renewcommand{\thefigure}{S\arabic{figure}}
\renewcommand{\bibnumfmt}[1]{[S#1]}
\renewcommand{\citenumfont}[1]{S#1}

\section{Collective modes of the bcc crystal}

The excitations of the CDW order are obtained by considering small fluctuations, $\eta_{\BG}(x,t)$, about the mean field CDW solution,
\begin{equation}
\delta \rho(\Bx,t) = \sum_{\BG}\big(\delta \bar{\rho}+\eta\ns_{\BG}(x,t)\big)e^{i \BG \cdot \Bx}
\end{equation}
where the sum is over the twelve primary Bragg vectors of the bcc lattice.  The functions $\eta\ns_\BG(\Bx,t)$ are taken to be slowly varying in space relative to the reciprocal lattice vectors $|\BG|$.

Expanding the GL Lagrangian in Eq.(3) of the main text to quadratic order in                                                                                                                                                                                                                                                                                                                                                                                    the fluctuations yields the harmonic action,
\begin{equation}
\mathcal{S}_{\text {H}}=\sum_{\BG,\BG'}\int \! { d^3\! q\,d\omega\over(2\pi)^4}\>
\eta\ns_\BG(\Bq,\omega)\, \mathcal{A}\ns_{\BG,\BG'}(q,\omega) \, \eta\ns_{\BG'}(-\Bq,-\omega).
\end{equation} 
In the above equation the kernel matrix is given by:
\begin{eqnarray}
\label{eq:har_GGp}
\mathcal{A}_{\BG,\BG'}(q,\omega)&=&\big(\gamma^{-1}\omega^2-\tilde{\chi}^{-1}(\BG+\Bq)\big)\,\delta_{\BG,-\BG'} -B\,\delta \bar{\rho}\!\!\! \sum_{\BG_1,\BG_2,\BG_3}\!\!\! \delta\left(\BG_1+\BG_2+\BG_3\right)\,\sum_{i<j}^3 \delta_{\BG,\BG_i} \delta_{\BG',\BG_j} \\
&&\qquad+ C\,\delta \bar{\rho}^2\!\!\! \sum_{\substack{\BG_1,\BG_2\\ \BG_3,\BG_4}}\!\!\! \delta\left(\BG_1+\BG_2+\BG_3+\BG_4\right)\,\sum_{i<j}^4 \delta_{\BG,\BG_i} \delta_{\BG',\BG_j} \nonumber
\end{eqnarray} 
where $\tilde{\chi}^{-1}({\BG+\Bq})=\half R+\half v^2\left((\BG+\Bq)^2-G^2\right)^2$.

The density $\delta \rho(x,t)$ is real.  Therefore, fields related by inversion are complex conjugate pairs $\eta_{-\BG}(x,t)=\eta^*_\BG(x,t)$. 
A convenient choice for the physical degrees of freedom for a given pair of Bragg vectors $\left\{\BG,-\BG\right\}$ is $\eta^e_\BG(x,t)=\Real\left\{\eta_\BG(x,t)\right\}$ and $\eta^o_\BG(x,t)=\Imag\left\{\eta_\BG(x,t)\right\}$. The field $\eta^e_\BG(x,t)\,(\eta^o_\BG(x,t))$ is even (odd) under reflection $\BG\to-\BG$.  
In terms of these degrees of freedom the bilinear has the following form,
\begin{equation}
\label{eq:har_sa}
\mathcal{S}_{\text {H}}=
\sum_{\substack{\BG,\BG'>0\\ r,r'=\cbra{e,o}}}\int \! { d^3\! q\,d\omega\over(2\pi)^4}\> \eta_\BG^r(\Bq,\omega)\, \mathcal{B}_{\BG,\BG'}^{r,r'}(\Bq,\omega) \, \eta_{\BG'}^{r'}(-\Bq,-\omega),
\end{equation} 
where the index $r=\cbra{e,o}$ is used to label the even and odd modes. In order not to double-count degrees of freedom, we arbitrarily choose one vector $\BG$ from each pair  $\left\{\BG,-\BG\right\}$, and call it ``positive''.

The last two terms in Eq.~\eqref{eq:har_GGp} are symmetric under inversion $\BG \to-\BG$ thus they do not couple between the even and odd modes. On the other hand,  at finite momentum $\Bq\ne 0$ the susceptibility kernel $\tilde{\chi}^{-1}(\BG+\Bq)$ is not symmetric under inversion, and hence the kernel matrix $\mathcal{B}_{\BG,\BG'}^{r,r'}(\Bq,\omega)$ has nonvanishing off diagonal elements between even and odd modes. 

The kernel matrix $\mathcal{B}$ is Hermitian, $\mathcal{B}_{\BG,\BG'}^{r,r'}=\left(\mathcal{B}_{\BG',\BG}^{r',r}\right)^*$. Therefore it can be diagonalized by a unitary transformation 
\begin{eqnarray}
\eta_{\BG}^r=\sum_\alpha U_{\cbra{\BG,r},\alpha}\xi_\alpha\,.\label{eq:unitaryNormal}
\end{eqnarray}
This yields the resonance frequencies $\omega_\alpha(q)$ and their corresponding normal modes $\xi_\alpha(q,\omega)$.

\subsection{Excitation spectrum at $q=0$}

\begin{figure}
	\includegraphics{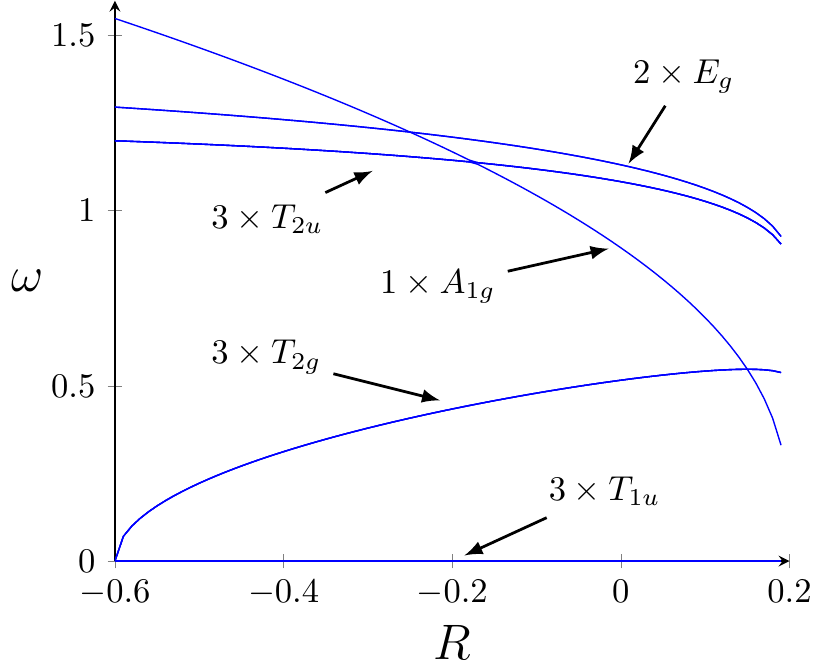}
	\caption{Excitation spectrum at zero momentum $q=0$ as a function of the GL parameter $R$.   The remaining GL parameters are taken to be $B=C=1$ and $v=1.5$.  This can always be done by expressing density, space, time, and energy in appropriate units.  Note that at $R^*\approx 0.15$ the lowest energy gapped modes exchange order. The modes are classified according to the irreducible representations of the octahedral group.}
	\label{fig:spect_qzero}
\end{figure}

We depict the zero relative momentum, $q=0$, excitation spectrum as a function of the GL parameter $R$ in Fig.~\ref{fig:spect_qzero}. The s-wave mode is lower (higher) in energy than d-wave mode for large (small) values of $R$. 

\begin{figure}
  \centering
  \subfloat[$s$]{\includegraphics[width=0.18\textwidth]{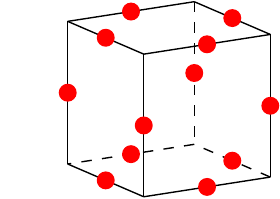}}
  \subfloat[$p_x$]{\includegraphics[width=0.18\textwidth]{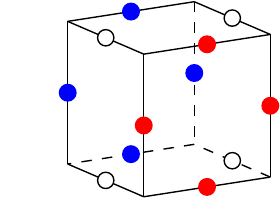}}
  \subfloat[$p_y$]{\includegraphics[width=0.18\textwidth]{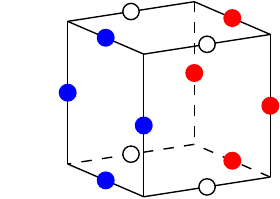}}
  \subfloat[$p_z$]{\includegraphics[width=0.18\textwidth]{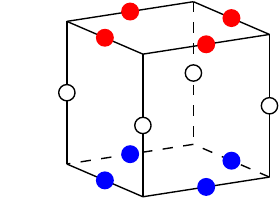}}\\

  \vskip 4mm
  
  \subfloat[$d_{xy}$]{\includegraphics[width=0.18\textwidth]{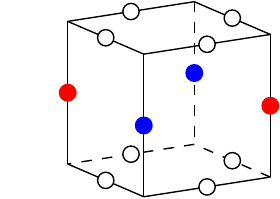}}
  \subfloat[$d_{yz}$]{\includegraphics[width=0.18\textwidth]{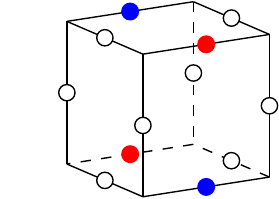}}
  \subfloat[$d_{xz}$]{\includegraphics[width=0.18\textwidth]{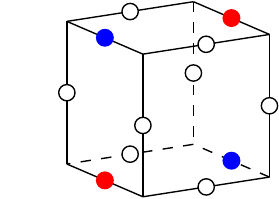}}
  \subfloat[$d_{x^2-y^2}$]{\includegraphics[width=0.18\textwidth]{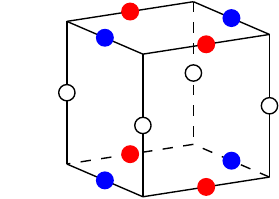}}\\  

  \vskip 4mm
  
  \subfloat[$d_{x^2}$]{\includegraphics[width=0.18\textwidth]{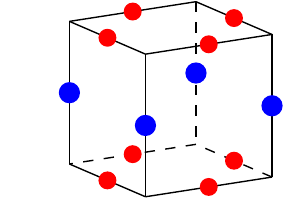}}
  \subfloat[$f_{z(x^2-y^2)}$]{\includegraphics[width=0.18\textwidth]{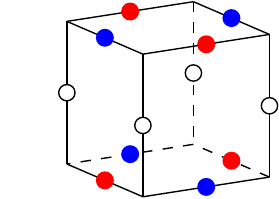}}
  \subfloat[$f_{y(z^2-x^2)}$]{\includegraphics[width=0.18\textwidth]{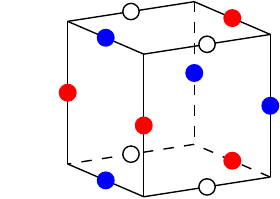}}
  \subfloat[$f_{x(z^2-y^2)}$]{\includegraphics[width=0.18\textwidth]{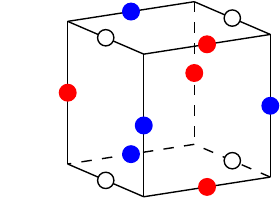}}\\  

\caption{Reciprocal space structure of the normal modes $\xi_\alpha$. Blue (red) dots correspond to positive (negative) amplitude $\eta_{\BG}$ for the different primary Bragg vectors $\BG$.  For odd-parity modes ($p$ and $f$-wave), $\eta_\BG$ are multiplied by an extra factor of $i$ to make them purely imaginary.  Modes are labelled by spherical harmonics.  In terms of irreps of the octahedral group $O_h$, $s\to A_{1g }$, $\{p_x,p_y,p_z\}\to T_{1u}$, $\{d_{xy},d_{yz},d_{xz}\}\to T_{2g}$, $\{d_{x^y-y^2},d_{z^2}\}\to E_g$, and $\{f_{z(x^2-y^2)},f_{y(z^2-x^2)},f_{x(z^2-y^2)}\} \to T_{2u}$.}
\label{fig:r_space}
\end{figure}

We illustrate the symmetry properties of the normal modes In Fig.~\ref{fig:r_space}, where we depict the zero momentum reciprocal space structure of the different modes.

\section{Dynamic structure factor}

The zero temperature dynamic structure factor at momentum $p=G+q$ is given by,
\begin{equation}
S(\BG+\Bq,\omega>0)=\Imag\left\{\avbra{\eta_\BG(\Bq,\omega)\,\eta_{-\BG}(-\Bq,\omega)}\right\},
\label{eq:sq}
\end{equation}
where expectation values are defined as, $\avbra{\mathcal{O}}= N \int \prod_{\BG>0,r}\mathcal{D} \eta_{\BG,r}\, e^{iS_{\text {H}}} $, where $N$ is a normalization constant.
To compute the dynamical structure factor we write Eq.~\eqref{eq:sq} in terms of the even and odd modes,
\begin{equation}
S(\BG+\Bq,\omega)=\Imag\left\{\avbra{\left|\eta_\BG^{e}(\Bq,\omega)+i \eta_{\BG}^{o}(\Bq,\omega)\right|^2}\right\} \nonumber
\end{equation}
and express them as a linear combination of the normal modes $\eta_{\BG}^r=\sum_{\alpha}U_{\left\{\BG,r\right\},\alpha}\,\xi_\alpha$
\begin{equation}
S(\BG+\Bq,\omega)=\Imag\left\{\bigg\langle\left|\sum_{\alpha}\left(U_{\left\{\BG,e\right\},\alpha}+iU_{\left\{\BG,o\right\},\alpha}\right)\xi_\alpha(\Bq,\omega)\right|^2\bigg\rangle\right\}
\end{equation}
We then use the normal modes' single particle Green's function, $\avbra{\xi_\alpha(\Bq,\omega_i)\,\xi_\beta(-\Bq,-\omega)}=
\gamma \delta\ns_{\alpha\beta}\big/\big[(\omega+i\epsilon)^2-\omega_{\alpha}^2(\Bq)\big]$ to obtain, at zero temperature,
\begin{equation}
S(\BG+q,\omega>0)=\sum_{\alpha}{\gamma M_{\BG,\alpha} \over 2\omega\ns_\alpha(\Bq)}\,\delta\big(\omega-\omega_\alpha(\Bq)\big)
\end{equation}
where the matrix element $M_{\BG,\alpha}$ is given in terms of the unitary matrix Eq.~\eqref{eq:unitaryNormal} by  
\begin{eqnarray}
M_{\BG,\alpha}=
\left|
U_{\left\{\BG,e\right\},\alpha}
+iU_{\left\{\BG,o\right\},\alpha}
\right|^2\,.
\end{eqnarray}
Note that $\sum_\alpha M_{\BG,\alpha}=2$, as follows from orthogonality of the columns of a unitary matrix.  Therefore,  the spectral function is constrained to satisfy the sum rule
\begin{eqnarray}
\int\limits_{-\infty}^\infty \!\!\!d\omega\, \omega\, S(\BG+\Bq,\omega)=\gamma
\end{eqnarray}

 \begin{figure}[b]
\centering
  \subfloat[Longitudinal]{\includegraphics[width=0.45\textwidth]{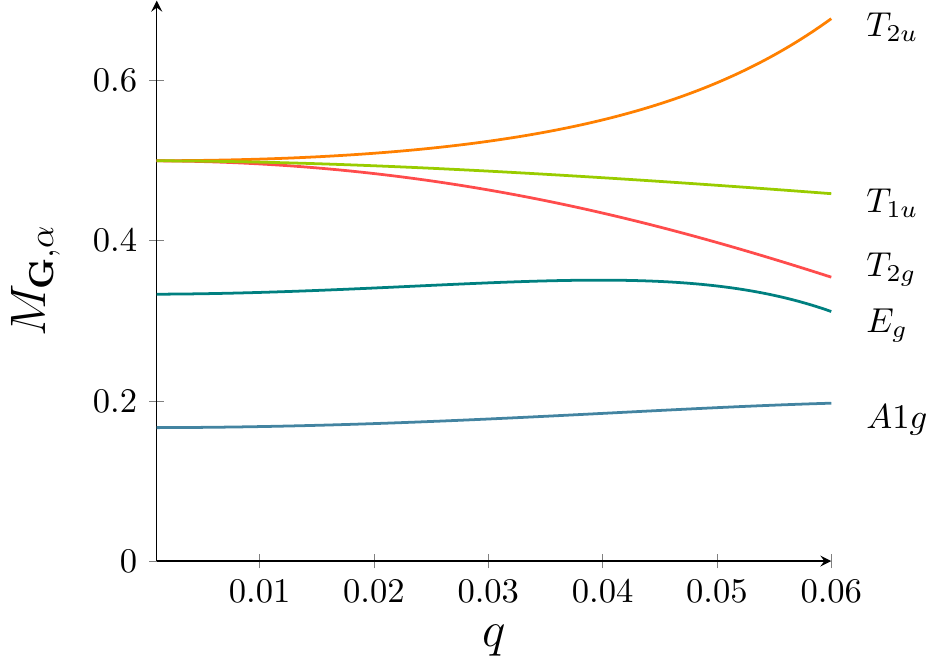}\label{sup_lon}}
  \subfloat[Transverse]{\includegraphics[width=0.45\textwidth]{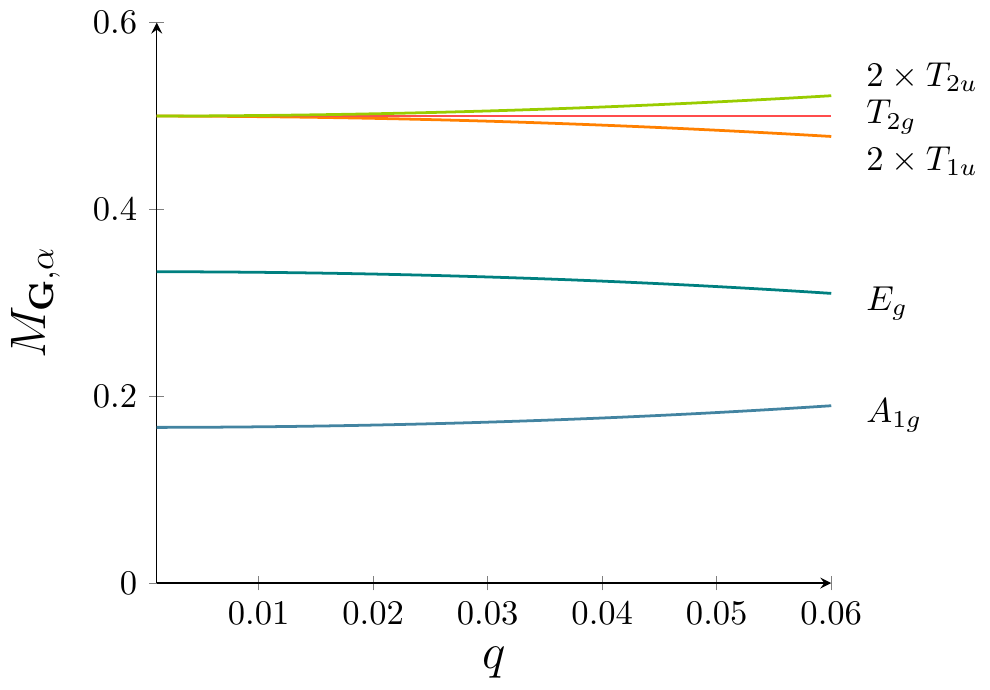}\label{sup_tra}}
\caption{$M_{\BG,\alpha}$ near the Bragg vector $\BG=\left\{1,1,0\right\}$ in the (a) longitudinal $\Bq=\left\{1,1,0\right\}$ and (b) transverse $\Bq=\left\{0,0,1\right\}$ directions.}
\label{fig:sf}
\end{figure}

The spectral weight associated with the matrix element $M_{\BG,\alpha}$ is depicted in Fig.~\ref{fig:sf} both for the longitudinal (Fig. \ref{sup_lon}) and transverse (Fig. \ref{sup_tra}) directions. The spectral weight of degenerate modes is summed over. In the longitudinal (transverse) direction only five (seven) modes out of the twelve modes have a non zero spectral weight and hence are visible in neutron scattering experiments. In the transverse direction the neutron scattering active modes consist of two pairs of degenerate modes one with $p$-wave symmetry and the other with $f$-wave symmetry. 

The result can be generalized to finite temperature following the fluctuation dissipation theorem
\begin{equation}
S(\BG+q,\omega,T)=\sum_{\alpha}{\gamma M_{\BG,\alpha}(\Bq)\over 2\omega_{\alpha}(\Bq)}\Big[\big(1+n(\omega_\alpha)\big)\,\delta\big(\omega-\omega_\alpha(\Bq)\big)+
n(\omega_\alpha)\,\delta\big(\omega+\omega_\alpha(\Bq)\big)\Big],
\end{equation}
where $n(\epsilon)=1\big/[\exp(\beta \epsilon)-1]$ is the Bose function.

\section{Extracting the energy scale of the amplitude mode from QMC data}

\begin{figure}[h]
			\includegraphics[scale=.9]{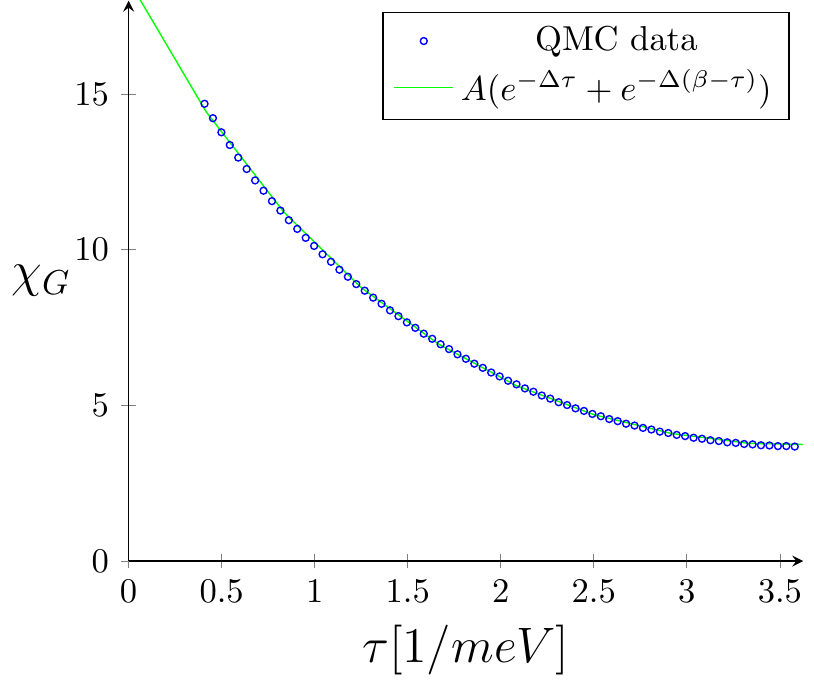}
	\caption{ Charge susceptibility at the Bragg momentum as a function of imaginary time (blue circles). The constant order parameter $\chi_0$ was subtracted. Numerical fit to an exponential decay  (solid green line). }
	\label{fig:expdecay}
\end{figure}

In Fig.~\ref{fig:expdecay} we depict the charge susceptibility evaluated at the Bragg momentum as a function of imaginary time after subtracting the order parameter $\chi_0$. The energy of the mode is extracted by a numerical fit  of the QMC data to an exponentially decaying function (see main text). We find very good agreement for $\Delta=0.6(1)$ meV. 

The simulation is carried on a finite system, and as such, it does not undergo true spontaneous symmetry breaking.  Could this be the source of the gapped excitation seen in our numerical simulation?  In particular, the finite system size gives rise to an Anderson tower of states, corresponding to center of mass motion of the entire system.  For $N$ helium atoms, each with mass $m$, this yields a mode at the Bragg vector $\BG$ of energy $\hbar^2 \BG^2/(2N m)$.  For our case $N=2000$ , this equals 2.43 meV$/2000=1.21\,\mu$eV, which is two orders of magnitude smaller than $\Delta$.  Hence, center of mass motion cannot account for the gapped excitation.

\section{Two phonon contribution to spectral function}
\begin{figure}[h]
\includegraphics[scale=1.3]{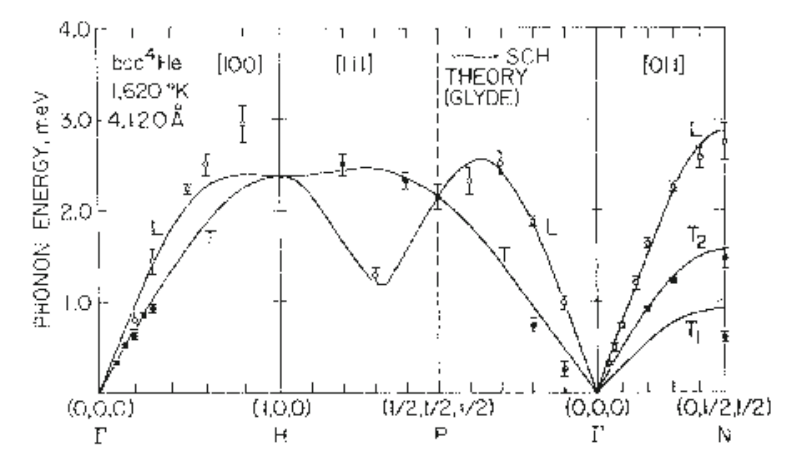}
\caption{Comparison of experimental and theoretical phonon dispersions in bcc ${}^4$He at molar volume $v_0=21.0\,{\rm cm}^3$.  (Fig. 9 from
Ref. \cite{PhysRevA.5.1537}.}
\label{fig:osgood}
\end{figure}

Here we consider the possibility that conventional two phonon processes could be responsible for the experimentally observed peak in $S(\BG,\omega)$
\cite{Bert}.  We analyze a harmonic theory including up to third-neighbor interactions which adequately reproduces key features of the neutron data
\cite{PhysRevA.5.1537} and previous self-consistent harmonic theories of the phonon spectra \cite{Glyde1971} shown in Fig. \ref{fig:osgood}.  

The content of the harmonic theory for excitations of a Bravais lattice is embodied in the dynamical matrix,
\begin{equation}
\sPhi^{\mu\nu}(\Bk) = {\sum_\BR}' (1-\cos\Bk\cdot\BR)\,{\pz^2 v(\BR)\over\pz R^\mu\,\pz R^\nu}\quad,
\end{equation}
where the sum is over all finite length direct lattice vectors $\BR$, and $v(\BR)$ is the interatomic potential.  The eigenvalue equation for the
$a^{\rm th}$ acoustic phonon branch is $m\omega_a^2(\Bk)\,e^a_\mu(\Bk)=\sPhi^{\mu\nu}(\Bk)\,e^a_\nu(\Bk)$, where the polarization
vectors satisfy completeness $\sum_a e^{a*}_{\mu}(\Bk)\,e^a_{\nu}(\Bk)=m^{-1}\,\delta\ns_{\mu\nu}$ and orthogonality
$\sum_{\mu} e^{a*}_{\mu}(\Bk)\,e^b_{\mu}(\Bk)=m^{-1}\delta^{ab}$.  Again, $m$ is the atomic mass.  Assuming a central potential $v(R)$, we have
\begin{equation}
{\pz^2 v(R) \over\pz R^\mu \, \pz R^\nu} = \big(\delta^{\mu\nu}-\HR^\mu\HR^\nu\big)\,
{v'(R)\over R} + \HR^\mu \HR^\nu \, v''(R) \ ,
\label{vpvpp}
\end{equation}
where $\HR^\mu=R^\mu/|\BR|$.  Thus, a phenomenological model which includes interactions up to $p^{\rm th}$ nearest neighbors is specified
by $2p$ parameters, which are the values of $v(R)$ and $v'(R)$ for the first $p$ separation vectors.

We define the angles $(\theta\ns_1,\theta\ns_2,\theta\ns_3,\theta\ns_4)$ by $\Bk\equiv\sum_{j=1}^3 \theta\ns_j\,\Bb\ns_j/2\pi$,
where $\Bb\ns_j$ is an elementary reciprocal lattice vector, and $\theta\ns_4\equiv\theta\ns_1+\theta\ns_2+\theta\ns_3$.
The contribution from the eight first neighbors may then be written as
\begin{align}
\sPhi^{(1)}_{\alpha\beta}(\Bk)&=\fourth A\,(4-c\ns_1-c\ns_2-c\ns_3-c\ns_4)\,{\mathbb I}\bvph\\
&\hskip0.5in + 
\fourth\,B\begin{pmatrix} 0 & c\ns_3+c\ns_4-c\ns_1-c\ns_2 & c\ns_2+c\ns_4-c\ns_1-c\ns_3 \\ c\ns_3+c\ns_4-c\ns_1-c\ns_2 & 0 & c\ns_1 + c\ns_4 - c\ns_2 - c\ns_3 \\
c\ns_2+c\ns_4-c\ns_1-c\ns_3 &  c\ns_1 + c\ns_4 - c\ns_2 - c\ns_3 & 0 \end{pmatrix}\nonumber
\end{align}
where $c\ns_j=\cos\theta\ns_j$, $A=\frac{4}{3}\big[ 2a^{-1} v'(a) + v''(a)\big]$, $B=\frac{4}{3}\big[ a^{-1} v'(a) -2 v''(a)\big]$, and $a$ is the nearest neighbor
separation.

The contribution to the dynamical matrix from the six second neighbors is diagonal:
\begin{align}
\sPhi^{(2)}_{\alpha\beta}(\Bk)&=\textsf{diag}\big(\Lambda\ns_1(\Bk)\,,\,\Lambda\ns_2(\Bk)\,,\,\Lambda\ns_3(\Bk)\big) \\
\Lambda\ns_1(\Bk)&=\half\,C\,(2-c\ns_{12}-c\ns_{13}) + \half\,D\,(1-c\ns_{23}) \nonumber \\
\Lambda\ns_2(\Bk)&= \half\,C\,(2-c\ns_{12}-c\ns_{23}) + \half\,D\,(1-c\ns_{13}) \nonumber \\
\Lambda\ns_3(\Bk)&=  \half\,C\,(2-c\ns_{13}-c\ns_{23}) + \half\,D\,(1-c\ns_{12})\quad,\nonumber
\end{align}
where $c\ns_{12}=\cos(\theta\ns_1+\theta\ns_2)$, $c\ns_{13}=\cos(\theta\ns_1+\theta\ns_3)$, $c\ns_{23}=\cos(\theta\ns_2+\theta\ns_3)$,
$d=2a/\sqrt{3}$ is the second neighbor separation, $C=8v'(d)/d$, and $D=4 v''(d)$.

Finally, there are eight third neighbors, contributing
\begin{align}
\sPhi^{(3)}_{\alpha\beta}(\Bk)&=\fourth E\,(4-\ctil\ns_1-\ctil\ns_2-\ctil\ns_3-\ctil\ns_4)\,{\mathbb I}\bvph\\
&\hskip0.5in + 
\fourth\,F\begin{pmatrix} 0 & \ctil\ns_3+\ctil\ns_4-\ctil\ns_1-\ctil\ns_2 & \ctil\ns_2+\ctil\ns_4-\ctil\ns_1-\ctil\ns_3 \\
\ctil\ns_3+\ctil\ns_4-\ctil\ns_1-\ctil\ns_2 & 0 & \ctil\ns_1 + \ctil\ns_4 - \ctil\ns_2 - \ctil\ns_3 \\
\ctil\ns_2+\ctil\ns_4-\ctil\ns_1-\ctil\ns_3 &  \ctil\ns_1 + \ctil\ns_4 - \ctil\ns_2 - \ctil\ns_3 & 0 \end{pmatrix}\quad,\nonumber
\end{align}
where $\ctil\ns_j=\cos(2\theta\ns_j)$, $E=\frac{4}{3}\big[ a^{-1} v'(2a) + v''(2a)\big]$, and $F=\frac{4}{3}\big[ \half a^{-1} v'(2a) -2 v''(2a)\big]$.

The values of $\Btheta=(\theta\ns_1,\theta\ns_2,\theta\ns_3)$ at the high symmetry points in the Brillouin zone $\Gamma$, N, P, and H are
$\Btheta\ns_\Gamma=(0,0,0)$, $\Btheta\ns_\RN=(\pi,0,0)$, $\Btheta\ns_\RP=(\frac{\pi}{2},\frac{\pi}{2},\frac{\pi}{2})$, and $\Btheta\ns_\RH=(-\pi,\pi,\pi)$.
The phonon frequencies at these high symmetry points are then
\begin{equation}
\begin{split}
\Gamma:&\quad M\omega_1^2=M  \omega_2^2=M\omega_3^2=0 \\
\RN:&\quad M\omega_1^2= A-B+C+D \quad,\quad M\omega_2^2= A+2C \quad,\quad M\omega_3^2=A+B+C+D \bvph\\
\RP:&\quad M\omega_1^2=M\omega_2^2=M\omega_3^2=A+2C+D+E \\
\RH:&\quad M\omega_1^2=M\omega_2^2=M\omega_3^2=2A\quad . \bvph
\end{split}
\end{equation}

\begin{figure}[t]
\includegraphics[scale=0.5]{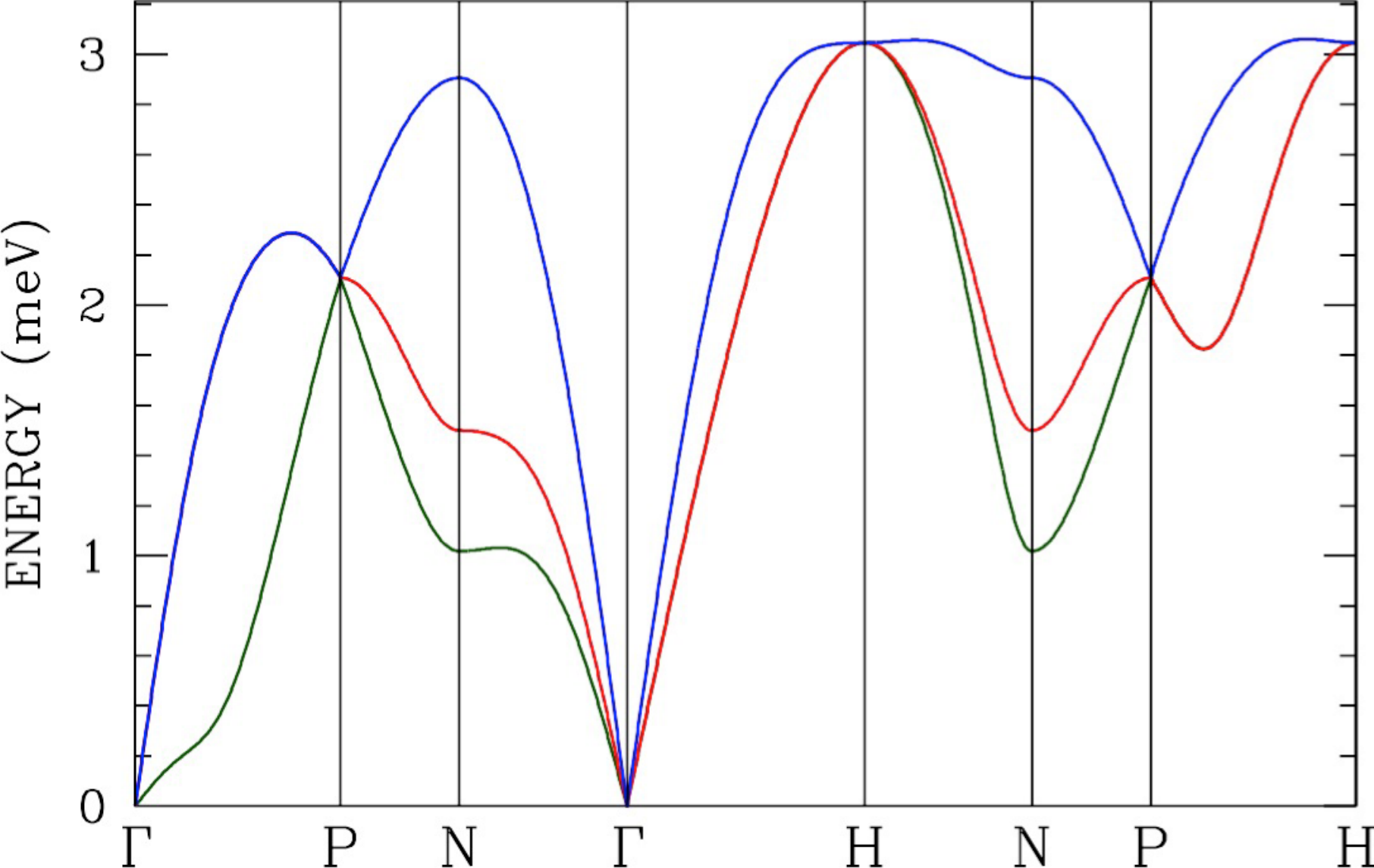}
\caption{Phonon band structure within the third nearest neighbor model, with $A=51.6$, $B=41.2$, $C=-13.3$, $D=14.4$, $E=5.0$, and $F=1.0$,
all in units of ${\rm K}/{\rm\AA}^2$. }
\label{fig:pbs}
\end{figure}

At the high symmetry points, the theoretical values of the nonzero phonon frequencies, from Fig. \ref{fig:osgood}, are
\begin{equation*}
\omega\ns_\RP\simeq 2.1\,{\rm meV}\ ,\  \omega\ns_\RH\simeq 2.4\,{\rm meV}\ ,\  \omega\ns_{\RN,1}\simeq1\,{\rm meV}\ ,\ 
\omega\ns_{\RN,2}\simeq 1.5\,{\rm meV}\ ,\  \omega\ns_{\RN,3}\simeq 2.9\,{\rm meV}\quad.
\end{equation*}
We find that these may be accurately fit by a second neighbor model, with $A=31.6$, $B=41.2$, $C=-3.28$, and $D=24.4$, all in units of
${\rm K}/{\rm\AA}^2$ (with $E=F=0$).  Since this entails solving five equations in four unknowns, there is a dependence relation,
$\omega^2_{\RN,1}+\omega^2_{\RN,2}+\omega^2_{\RN,3}=2\omega^2_\RP+\half \omega^2_\RH$, which, remarkably, is well-satisfied by
the previous theoretical predictions based on self-consistent phonon theory, with the LHS of this relation yielding $11.66\,({\rm meV})^2$ and the RHS
$11.7\,({\rm meV})^2$.  However, we find this second neighbor model is unstable, giving imaginary phonon frequencies along a portion of the
$\Gamma\RP$ segment.  By including third neighbor terms $E$ and $F$, the problem of matching the five nonzero high symmetry phonon frequencies
becomes underdetermined, and we may use the resulting freedom to fix the instability problem.
We find that by choosing
\begin{equation*}
A=51.6\quad,\quad    B=41.2\quad,\quad    C=-13.3\quad,\quad
D=14.4\quad,\quad E=5.0\quad,\quad F=1.0\quad,
\end{equation*}
all in units of $\RK/{\rm\AA}^2$, we obtain the same phonon frequencies at the $\Gamma$, N, and P points, at the expense of shifting the H point
value from 2.4 meV up to about 3 meV.  In fact, the experimental value close to H, reported in Ref. \cite{PhysRevA.5.1537}, is itself very close to 3 meV. 

We now compute the dynamic structure factor $S(\Bq,\omega,T)$.  For a harmonic Bravais lattice, is given by
\begin{equation}
S(\Bq,\omega,T)=e^{-2W(\Bq,T)}\!\!\impi\!{dt\over 2\pi\hbar}\>\sum_\BR e^{-i\Bq\cdot\BR}\,e^{i\omega t}\,e^{\sGamma(\Bq,T;\BR,t)}\quad,
\label{DSFeqn}
\end{equation}
where $N$ is the number of particles and $e^{-2W(\Bq,T)}$ is the Debye-Waller factor, with
\begin{equation}
2W(\Bq,T)=\!\int\limits_\HOmega\!{d^d\!k\over (2\pi)^d}\>\sum_{a=1}^3 {\hbar\,v\ns_0\over 2\,\omega\ns_a(\Bk)}\,\big|\Bq\cdot \Be^a(\Bk)\big|^2
\ctnh\Big(\frac{\hbar\omega\ns_a(\Bq)}{\kT}\Big)
\end{equation}
and
\begin{align}
\Gamma(\Bq,T;\BR,t)&=\!\int\limits_\HOmega\!{d^d\!k\over (2\pi)^d}\>\sum_{a=1}^3 {\hbar\,v\ns_0\over 2\,\omega\ns_a(\Bk)}\,\big|\Bq\cdot \Be^a(\Bk)\big|^2
\bigg\{ \Big(n\ns_a(\Bk,T) + 1\Big)\,e^{i\Bk\cdot\BR}\,e^{-i\omega\ns_a(\Bk)t}\\
&\hskip3.0in + n\ns_a(\Bk,T) \,e^{-i\Bk\cdot\BR}\,e^{i\omega\ns_a(\Bk)t}\bigg\}\quad.\nonumber
\end{align}
Here $v\ns_0$ is the unit cell volume, the integrals are over the first Brillouin zone $\HOmega$, and $n\ns_a(\Bk,T)$
is the thermal occupation of the $\sket{a,\Bk}$ phonon state.

\begin{figure}[t]
\includegraphics[scale=0.5]{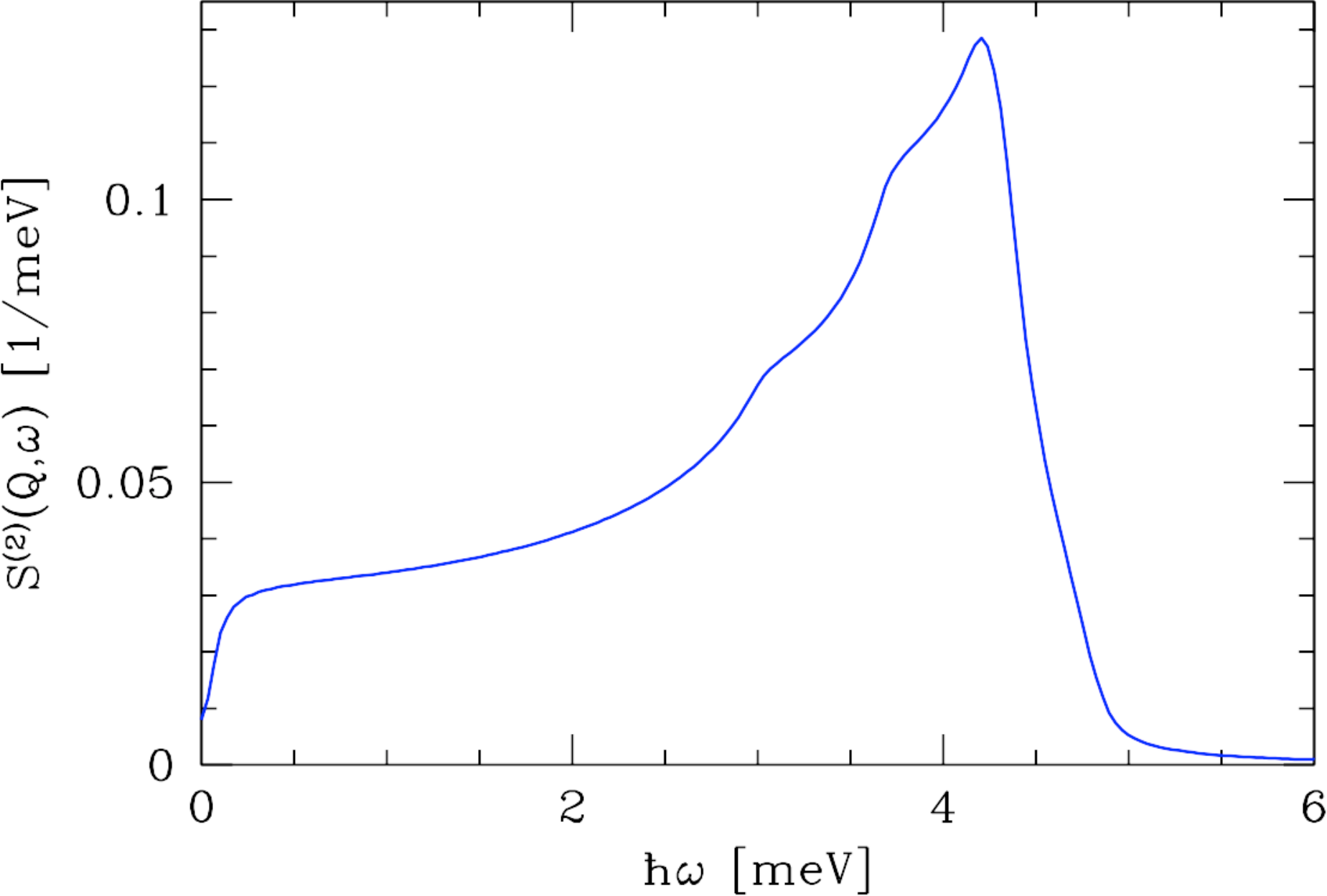}
\caption{Two phonon contribution $S^{(2)}(\BQ,\omega)$ to the dynamic structure factor, with $\BQ=\Bb_1$, for the third neighbor harmonic lattice model,
with parameters given in the caption to Fig. \ref{fig:pbs}, and molar volume $v_0=21.0\,{\rm cm}^3$.}
\label{fig:sqwh}
\end{figure}

Since the experiments are performed at $T=1.4\,$K, which is much lower than the phonon bandwidth, we will set $T=0$.  Note that while the bcc phase
is unstable at $T=0$ for physical interatomic potentials, there is no such problem for our third neighbor model.  Expanding the integrand in
Eqn. \ref{DSFeqn} in powers of $\sGamma(\Bq,T=0\,;\BR,t)$, we obtain a series
$S(\Bq,\omega)=S^{(0)}(\Bq,\omega)+S^{(1)}(\Bq,\omega)+S^{(2)}(\Bq,\omega)+\ldots\,$, where $S^{(r)}(\Bq,\omega)$ is the contribution from
$r$-phonon processes.  When $\Bq=\BG$ is a reciprocal lattice vector, the single phonon contribution vanishes at finite frequency.  The two phonon
contribution is
\begin{equation}
S^{(2)}(\Bq,\omega)=\half\hbar v\ns_0\,e^{-2W(\Bq)}\!\!\int\limits_\HOmega\!\!{d^d\!k\over (2\pi)^d}\sum_{a,b=1}^3
{\big|\Bq\cdot\Be^a(\Bk)\big|^2\over 2\,\omega\ns_a(\Bk)}\,
{\big|\Bq\cdot\Be^b(\Bq-\Bk)\big|^2\over 2\,\omega\ns_b(\Bq-\Bk)}\,\delta\big(\omega-\omega\ns_a(\Bk)-\omega\ns_b(\Bq-\Bk)\big)\quad.
\label{SSF}
\end{equation}
We evaluate the above expression within our third neighbor model by discretizing the Brillouin zone integral over a $120\times120\times120$ mesh,
and approximating the delta function in frequency by a Lorentzian $\delta(\nu)\approx \gamma/ \pi(\nu^2+\gamma^2)$, with $\gamma=0.06\,$meV.
The results are shown in Fig. \ref{fig:sqwh} for a molar volume $v\ns_0=21.0\,{\rm cm}^3$.  We find that the peak occurs at a frequency of approximately
$4.3\,$meV, which may be roughly understood in terms of the phonon spectrum in Fig. \ref{fig:pbs}.  The frequency of this peak is substantially greater than that for
the ``new excitation" observed by Markovich {\it et al.\/} \cite{Markovich}, which lies at $1.2\,$meV.  Our harmonic lattice third neighbor
model gives a variance in the atomic positions of $\langle {\mib u}^2\rangle= 0.90\,{\rm\AA}^2$, which is roughly consistent with the value $1.09\,{\rm\AA}^2$
given in Table 3.1 of Ref. \cite{glyde1994}.  We find a Debye-Waller factor of $2W(\BQ)=1.40$.  Higher order phonon processes will yield features at
higher frequencies still.  In this context it is worth emphasizing that the first moment of the $r$-phonon $S^{(r)}(\Bq,\omega)$ is
given by
\begin{equation}
I\ns_r(\Bq)\equiv\!\!\impi\!d\omega\,\omega\,S^{(r)}(\Bq,\omega)={\Bq^2\over 2m}\times{\big[2W(\Bq)\big]^{r-1}\over (r-1)!}\,e^{-2W(\Bq)}\quad,
\end{equation}
with $I\ns_0(\Bq)=0$ since $S^{(0)}(\Bq,\omega)\propto\delta(\omega)$.  Summing over all $r$, the first moment of $S(\Bq,\omega)$ itself is then
$\Bq^2/2m$, which reflects the $f$-sum rule.  Setting $\Bq=\BQ=\Bb\ns_1$, we obtain $\BQ^2/2m=2.43\,{\rm meV}/\hbar^2$, of which $I\ns_2(\BQ)=0.838\,{\rm meV}/\hbar^2$
is from two-phonon processes.  When $W$ is large, one finds that $I\ns_r$ is maximized for $r=r^*\approx 1+2W$.  Thus, beyond $r=2$, the magnitude of the contributions
to $S(\BQ,\omega)$ will decrease, and any additional prominences will lie at higher frequencies.  In addition, numerical integration of the QMC results in Fig. \ref{fig:qmc}
yields a first moment of $2.6\,{\rm meV}/\hbar^2$, which is only 7\% off from the sum rule. Both the location as well as the height of the QMC peak in $S(\BQ,\omega)$ are
substantially different than what the harmonic lattice calculation predicts.  Therefore it is very difficult for us to see how the observed experimental
peak at $\omega\approx 1.2\,$meV can be due to any multi-phonon processes.

\end{document}